# Aero-thermal analysis of a laminar separation bubble subjected to varying free-stream turbulence: Large Eddy Simulation

Sonalika Srivastava, and S. Sarkar [a]

*Department of Mechanical Engineering, Indian Institute of Technology Kanpur, Kanpur, 208016, Uttar Pradesh, India*

A quantitative analysis illustrating salient features of a Laminar Separation Bubble (LSB), its transition forming coherent structures, and associated heat transfer has been performed on a flat plate for varying free stream turbulence (*fst*) between 1.2% to 10.3%. A well-resolved Large Eddy Simulation (LES) developed in-house is used for the purpose. Flow separation has been induced by imposing an adverse pressure gradient on the upper boundary of a Cartesian domain. Isotropic perturbations are introduced at the inlet to mimic grid turbulence. With an increase of *fst*, an upstream shift in the mean reattachment point has been observed while the onset of separation remains almost invariant, shrinking the bubble length significantly. The transition of the shear layer is triggered by the Kelvin-Helmholtz (K-H) instability for *fst* of less than 3.3%, while Klebanoff modes (K-modes) dictate the flow transition at *fst* of greater than 6.5%. Further, a mixed mode, i.e., both K-H and K-modes, contribute to the flow transition at a moderate level of *fst*, lying between 3.3% and 6.5%. Thus, the roll-up of the shear layer appears in the second half of the bubble shedding large-scale vortices that keep their identity far downstream at low *fst* levels. On the contrary, the streamwise streaks via K-modes prior to the separation are found to interact with the LSB, resulting in an earlier breakdown of the shear layer with abundant small-scale vortices downstream at the moderate to high *fst* levels. However, higher surface-normal heat flux is associated with large-scale energetic coherent vortices.

## I. INTRODUCTION

In many engineering applications, the flow is prone to separation from the surface at low Reynolds number conditions, such as a low-pressure turbine, an aircraft wing, and small UAVs. The laminar boundary layer developing over such surfaces often separates under an adverse pressure gradient, undergoes transition due to enhanced receptivity to disturbances, becomes turbulent and finally reattaches to form a laminar separation bubble (LSB). Moreover, the turbine blades operate in an environment of high temperature and elevated free-stream turbulence, which complicates the features of such flows. The separated flow creates a dead air region in the first half, while large-scale vortex shedding and unsteadiness appear in the second-half, significantly influencing the local convective heat transfer from the surface. Therefore, thermal management in flows with separation is often critical and serves as a recurrent challenge to the research community. The available literature mainly focuses on aerodynamic behaviour, while significantly less attention has been given to heat transfer in separated flows in turbine-representative conditions. Further, the heat flux over blade surfaces in the presence of an LSB subjected to varying *fst* need to be predicted reliably to ensure the durability of an engine, apart from its design activities.

---

[a] Author to whom correspondence should be addressed. Electronic mail: subra@iitk.ac.in.

Transition in the disturbance-free environment in an attached boundary layer is initiated through Tollmien-Schlichting (T-S) waves, while in low disturbance ambience, some low-frequency disturbances pierce into the boundary layer due to the receptivity mechanism.[1] These disturbances formed at low *fst* levels in an attached flow may amplify due to downstream flow separation.[2-4] The separated flow is inviscidly unstable to a broad range of frequencies, among which the most unstable modes lead to flow transition.[5-7] Many studies corroborated that the amplification of selective frequencies is attributed to the Kelvin-Helmholtz (K-H) instability of the shear layer.[8-15] Further, the transition of the shear layer is followed by periodic vortex shedding from the second-half of the bubble leading to the breakdown and reattachment of the shear layer with abundant small-scale eddies, augmenting mixing downstream.[16-18]

Several studies have been performed to understand the laminar-turbulent transition in convectively growing perturbations.[11, 19-21] The excitation of an LSB subjected to the controlled disturbance was studied through experiments and simulations. It was reported that the transition of a shear layer is usually initiated through oblique modes and often Lambda-induced vortex breakdown.[3,10,22] The two-dimensional disturbance waves inside an LES are the most amplified modes, while the few weak oblique disturbances could also experience amplification.[10,22] In brief, the transition process and the evolution of disturbance modes after separation are intricately linked to the wave frequency, wavenumber and initial amplitudes resulting from the receptivity process.[23] It is noteworthy that the amplitude of spanwise waves does not affect the growth rate of the most unstable frequency, while on the contrary spanwise waves dictate the breakdown process of K-H rolls in the second half of the separation bubble.[10,22]

In simulating the real path to turbulence, it is appropriate to assume *fst* as a most general disturbance influencing the transition of an LSB. Imposed inlet disturbances often result in low-frequency streamwise streaks via Klebanoff modes in an attached boundary layer due to the shear-filtering process.[19, 24] It was also reported via a DNS of the attached boundary layer subjected to the *fst* of 5.3% that these streamwise streaks tend to stabilize under the favourable pressure gradient. In contrast, the secondary instability of these low-speed streaks triggers a transition leading to turbulence downstream in an adverse pressure gradient.[25] DNS of Wissink and Rodi[26] can be considered a prefatory work to understand the effect of *fst* on an LSB formed on a flat plate through an adverse pressure gradient. The transition of the shear layer due to KH instability appears earlier, with the *fst* level of 1.5%, leading to a reduction in bubble size, which is also confirmed by several experiments and numerical studies.[4,27,28] The elongated low-frequency streamwise structures formed via Klebanoff modes trigger earlier transition before the separation, leading to a rapid breakdown and shrinking the separation bubble with increasing levels of *fst*.[14,24,27] An elevated *fst* level of 10% might eliminate the separation bubble.[29] The recent study by Hosseinverdi & Fasel[30] for varying *fst* levels of 0-3% depicts the domination of streamwise streaks in triggering transition for an elevated level of *fst*, while for lower *fst* level, it is initiated through classical K-H. Thus, the level of *fst* not only alters bubble topology but also alters the associated transition mechanism.

There exists a scalar-passive relationship between thermal and flow fields. If Reynolds analogy holds, then the heat transfer must be affected in the same manner as the flow field under the influence of *fst*. However, the Reynolds analogy ceases to hold for the separated flow. In the presence of an LSB, the flow downstream of separation slowly approaches a canonical layer. The flow and heat transfer behaviour in this developing region is dictated by the upstream LSB, which is controlled by the *fst* that decides its transition mechanism. In the last three decades, several efforts have been made to develop correlations for heat transfer as a function of free-stream turbulence in attached flows.[31-37] Higher heat transfer on the blade surfaces and the stagnation region is elucidated for varying *fst* via cascade experiments.[38-42] Furthermore, separated flows are known to induce unsteadiness leading to significant and rapid heat transfer variations with enhanced heat transport compared to attached flow.[43-45]

The work of Spalart and Strelets[18] is notable in understanding the thermal characteristics of an LSB via a DNS on a flat plate, although the study mainly focused on the thermal features in the recovery region. Calzada et al.[46] conducted a thermal analysis of separated flows in an LPT blade passage, indicating that the wall-normal component of the velocity plays an essential role in heat transfer at the wall, which is higher than attached flows. However, the transition mechanism and scale-dependent dynamics, which are vital attributes to relate the generation of turbulence with heat transfer, have not been addressed. Although the influence of *fst* on an LSB transition and flow structures is well understood, the corresponding effect on the heat transfer associated with a separated flow remains relatively unexplored. In brief, the separated flow leading to high unsteadiness, turbulence levels and enriched vortex dynamics are the key factors to explain the scales transport.

With the development of high-fidelity time-resolved numerical simulations, DNS and LES emerge as reliable tools to resolve these aero-thermal features. Previous works have shown the capability of LES to predict flow and heat transfer accurately for these complex flow conditions.[47-50] The present study is motivated by these ideas and attempts to understand the transition mechanism and its effects on heat transfer in a separated flow for varying *fst* levels through the high-fidelity LES. Here, the flow physics and thermal characteristics of an LSB are investigated numerically for varying *fst* ranging from 1.2 to 10.3%. An LSB over a flat plate was produced by suction-blowing on the upper boundary following Hosseinverdi & Fasel[30], and it was subjected to realistic grid turbulence. The validation of the flow features has been performed with the available experiment[51] and DNS[30]. The focus of the present study is to resolve the transition of an LSB, its coherent structures, turbulent statistics, relaxation after the reattachment and thermal transport, illustrating the effect of *fst*. The quantitative analysis performed here would create a knowledge database that might be helpful to improvise the activities on turbulence modelling for these types of flows.

## II. NUMERICAL DETAILS

The incompressible filtered mass, momentum and energy equations in Cartesian coordinates can be expressed as,

$$\frac{\partial \tilde{u}_i}{\partial x_i} = 0 \tag{1}$$

$$\frac{\partial \tilde{u}_i}{\partial t} + \frac{\partial}{\partial x_j}\left(\tilde{u}_j \tilde{u}_i\right) = -\frac{\partial \tilde{P}}{\partial x_i} + \frac{1}{Re}\nabla^2 \tilde{u}_i - \frac{\partial \tilde{\tau}_{ij}}{\partial x_j} \tag{2}$$

$$\frac{\partial \tilde{\theta}}{\partial t} + \frac{\partial}{\partial x_j}\left(\tilde{u}_j \tilde{\theta}\right) = \frac{1}{Re\,Pr}\nabla^2 \tilde{\theta} - \frac{\partial \tilde{h}_j}{\partial x_j} \tag{3}$$

Where, $\tilde{u}_i$ and $\tilde{\theta}$ represent the dimensionless filtered velocity and temperature field, $\tilde{P}$ is the pressure, $\tilde{\tau}_{ij}$ is sub-grid scale stress, $\tilde{h}_j$ is temperature flux, $Re$ is Reynolds number, and $Pr$ is the Prandtl number of the fluid. In cartesian coordinates, streamwise, wall-normal and spanwise directions are represented by $x$, $y$ and $z$, respectively and corresponding velocities are defined as $u$, $v$ and $w$. The length and velocity scales are normalized with respect to a reference length ($L_\infty$) and the free stream velocity ($U_\infty$). This reference length $L_\infty$, corresponds to the chord length of a NACA 643-618 aerofoil [30]. The temperature is scaled by the inlet temperature ($T_{in}$) and the wall temperature ($T_w$) to obtain the non-dimensional temperature as, $\theta = (T - T_w)/(T_{in} - T_w)$.

The computational domain and the imposed boundary conditions are depicted schematically in Fig. 1(a). In Fig. 1(b), the streamwise and wall-normal velocities are imposed at the inlet, where the Falkner-Skan equations are solved to replicate the experimental data for the aerofoil[30]. A thermal boundary layer profile is also specified at the inlet, as shown in Fig. 1(b), with a boundary layer thickness $\delta_T = 0.976\delta\,Pr^{-0.334}$, where, $\delta$ and $\delta_T$ are the velocity and thermal boundary layer thickness, respectively. Figure 1(c) depicts a suction-blowing profile for the v-velocity at the top boundary. This profile is imposed as a Dirichlet condition to create an adverse-pressure gradient over the flat plate that replicates the suction surface of a NACA 643-618 aerofoil. The other velocity components and temperature are set to zero ($u = w = \theta = 0$). No-slip, isothermally heated wall conditions are applied at the bottom boundary ($u = v = w = 0$, $\theta = 1$). A non-reflecting boundary condition is imposed at the exit plane to avoid any significant mass redistribution during the passage of vortices through this plane.

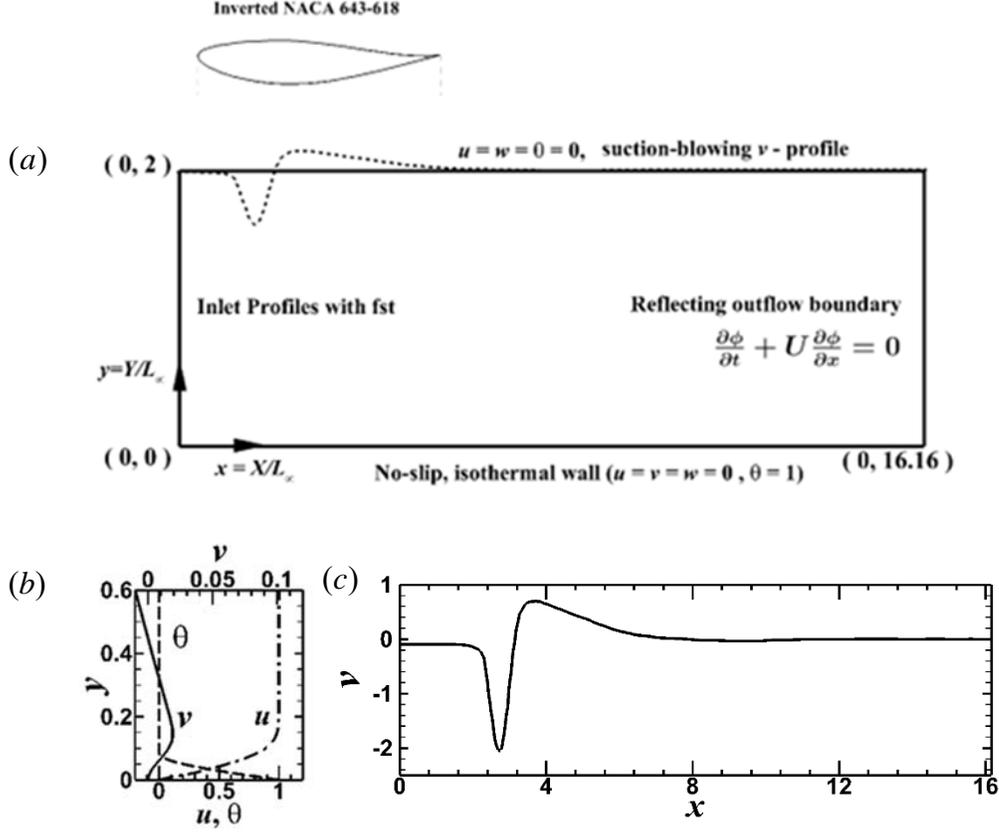

FIG. 1: (a) Computational domain with boundary conditions, (b) inlet velocity and temperature profiles and (c) suction-blowing profile of the $v$-velocity imposed at the top plane.

All the simulations are carried out using a $500 \times 240 \times 64$ grid. The grid spacing is uniform in the streamwise ($x$) and spanwise ($z$) directions. To resolve the inner layer characteristics, finer grids are used near the wall ($y \leq 0.6$), while the grids are slowly stretched in the wall-normal direction for $y > 0.6$. The near-wall resolutions are $\Delta x^+ = 4.81$ $\Delta y^+ = 0.32$ (for $y^+ \leq 9$) and $\Delta z^+ = 6.98$ near the reattachment for all the cases. The near-wall resolutions are evaluated following $\Delta^+ = \Delta \left( |C_{fr}/2| \right)^{1/2} Re$, where $C_{fr}$ is the skin-friction coefficient near the reattachment. The computational domain shown in Fig. 1(a) extends from the origin up to $16.16 L_\infty$ and $2 L_\infty$ in the streamwise and wall-normal directions, respectively, while the spanwise width is considered as $3L_\infty$. The flow Reynolds number is 3200, based on the reference length and free stream velocity at the inlet. The energy equation is solved for $Pr = 0.7$.

An in-house flow solver[12, 52-57], which is extensively validated in several studies for transitional and turbulent flows, is used here. The spatial discretization is second-order accurate using a symmetry-preserving central difference scheme, which is widely used in LES owing to its non-dissipative and conservative properties.[58, 59] The time advancement is explicit using the second-order Adams-Bashforth scheme[60], except for the pressure term, which is solved by a Projection method. The pressure equation is discrete Fourier transformed in the spanwise direction, in which the flow can be considered homogeneous. This results in a Helmholtz equation, which is solved by the BI-CG algorithm[61] using multi-grid acceleration in the other two directions. Such a Fourier technique yields the performance enhancement of the pressure solution compared to the three-dimensional Poisson equation by at least a factor of 5.[62] The solver has been parallelized using OPENMP. To model the SGS stress tensor, $\tilde{\tau}_{ij}$ a dynamic sub-grid scale model proposed by Germano et al.[63] and modified by Lilly[64] is used. The boundary layer is allowed to grow over the flat plate with the imposed boundary conditions. The solution is advanced with a time step of $\Delta t = 0.02$ in non-dimensional units that needs 10000 iterations for a flow pass. Seven flow passes have been allowed to develop turbulence and the separation bubble. Time-averaged statistics are taken for another ten flow passes. The simulations took around 5.6 CPU hours for a flow pass using 16 threads in an Intel Xeon Quadcore processor.

The LSB investigated here is subjected to the *fst*, which is modelled through velocity disturbances imposed at the inflow boundary of the computational domain. The imposed disturbance follows a spatial chessboard pattern in streamwise velocity and temporal fluctuations in wall-normal and spanwise velocities.[65] It can simulate streamwise vortices convecting at the local velocity while remaining divergence-free, mimicking grid turbulence. The equations used for generating perturbations in u, v and w components of velocity are prescribed as:

$$u' = \frac{4A}{\pi}\left(\cos(\theta_y) - \frac{1}{3}\cos(3\theta_y) + \frac{1}{5}\cos(5\theta_y) - \frac{1}{7}\cos(7\theta_y)\right) \\ \times \left(\cos(\theta_z) - \frac{1}{3}\cos(3\theta_z) + \frac{1}{5}\cos(5\theta_z) - \frac{1}{7}\cos(5\theta_z)\right) \quad (6)$$

$$v' = A\sin(\theta_z)\cos(\theta_y)\sin(2\pi tu'/l) \quad (7)$$

$$w' = A\cos(\theta_z)\sin(\theta_y)\sin(2\pi tu'/l) \quad (8)$$

Where $\theta_y = 2\pi y/l$ and $\theta_z = 2\pi z/l$, here, $l$ is the length scale of the perturbation at the inlet and $A$ represents the amplitude of perturbation. In the present simulation, $l$ is taken as inlet displacement thickness ($\delta_{in}^*$), and $A$ is specified as a fraction of inlet free stream velocity depending on the turbulence levels. The *fst* level is represented by the maximum turbulent intensity $Tu = \frac{1}{3}\left(u_{rms}^2 + v_{rms}^2 + w_{rms}^2\right)^{0.5}$, evaluated at the inlet plane.

## III. RESULTS AND DISCUSSION

In this section, aero-thermal characteristics of an LSB for a wide range of *fst*, i.e., $Tu = 1.2, 1.5, 3.3, 4.0, 5.5, 6.5, 10.3\%$, are discussed to illustrate the effect of the free-stream perturbations on the transition to turbulence and associated heat transfer behaviour. However, only those cases that distinctly change the fundamental flow physics, such as $Tu = 1.2, 3.3$ and $Tu = 10.3\%$, are discussed in detail. At first, the result for $Tu = 3.3\%$ is validated against the available experiment[51] and DNS[30]. Following this, a detailed analysis based on the instantaneous flow, mean quantities and turbulent statistics is carried out to acquire more profound insights into the problem.

### A. Validation Study

The time- and span-averaged streamwise velocity at various streamwise stations for $Tu = 3.3\%$ are compared with the experimental data[51] in Fig. 2. The figure shows reasonable agreement in the profiles with minor discrepancies that can be attributed to the uncertainties with the experiment. The variations of time-averaged skin friction coefficient for various *fst* levels are presented in Fig. 3, where the skin friction coefficient is defined as $C_f = \frac{2}{Re}\left(\partial u/\partial y\right)_{y=0}$. The distribution of the skin friction coefficient from a DNS[30] is also superimposed. The results from the present LES with the dynamic model at the turbulence intensity of 3.3% agree well with the DNS data at a turbulence level of 0.5%. Although the difference in turbulence intensity is not fully appreciated, the overall flow features and the progressive shift in $C_f$ with turbulence intensity are similar.

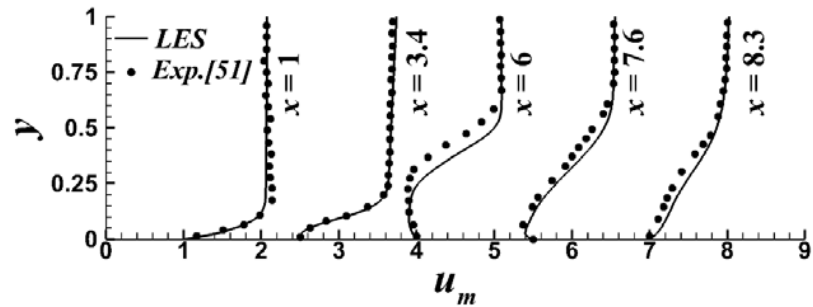

FIG. 2: Time and span averaged streamwise velocity compared with experimental data[51] at different streamwise stations.

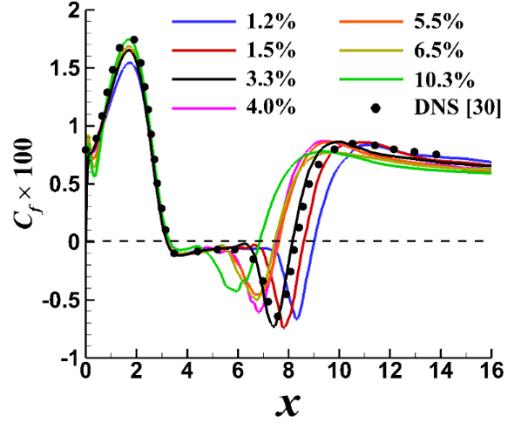

FIG. 3: Time and span averaged wall skin friction distribution in the streamwise direction at varying free stream turbulence.

TABLE I. Separation and reattachement points at various *fst* levels.

| $Tu_{in\%}$ | $x_s$ | $x_r$ | $l_b$ |
|---|---|---|---|
| 1.2 | 3.27 | 8.99 | 5.72 |
| 1.5 | 3.19 | 8.72 | 5.53 |
| 3.3 | 3.29 | 8.00 | 4.71 |
| 4.0 | 3.3 | 7.61 | 4.55 |
| 5.5 | 3.19 | 7.6 | 4.39 |
| 6.5 | 3.23 | 7.43 | 4.20 |
| 10.3 | 3.21 | 6.76 | 3.55 |

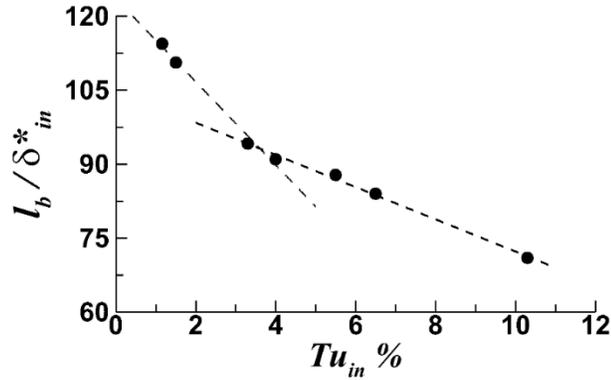

FIG. 4: Variation of bubble length (normalized by the inlet boundary layer thickness) turbulent intensity at the point of separation.

The separation and reattachment points of the shear layer, as evaluated by the zero crossings of the skin friction in Fig.3 are reported in Table I for different *fst* levels. The separation point ($x_s$) remains almost invariant with *fst*, while the reattachment location ($x_r$) shifts upstream with an increase of *fst*, reducing the mean bubble length ($l_b$). This upstream shift of reattachment point can be attributed to the early onset of transition with elevated *fst*.[30, 66] The plateau in $C_f$ distribution after the separation point ($x_s \approx 3$) corresponds to the dead air region, whereas the following larger negative skin friction indicates the reverse flow vortex. The streamwise extent of the dead air region reduces due to a faster transition with increasing *fst*, which will be discussed later. The variation of mean bubble length with inlet *fst* is presented in Fig.4. The slope of the curve shows that the mean bubble length decreases at a higher rate for lower levels of *fst* (1.0-4.0%) as compared to the rate with *fst* levels of 5.0-10.3%. Similar behaviour was observed by Hossienverdi and Fasel.[30]

## B. Vortex Dynamics, Transition and Three-Dimensional Motions

In this section, vortex dynamics, the transition of the shear layer, and the development of three-dimensional motions correlating to the thermal field are discussed at different *fst* levels. The instantaneous snapshots of the spanwise vorticity ($\omega_z$) superimposed with streamlines over a vortex-shedding period are depicted in Fig. 5, 6 and 7 at turbulence levels of 1.2%, 3.3% and 10.3%, respectively. The shedding period (T) is made dimensionless by the convective time scale ($L/U_\infty$). The contour of zero streamwise velocity ($u = 0$), denoted by a white line, demarcates the reverse flow region.

For $Tu = 1.2\%$ and $3.3\%$, the concentration of vorticity illustrates the formation of the shear layer, where the boundary layer separates at $x \approx 3.25$. The roll-up of the shear layer is apparent in the second half of the bubble, revealing the cat's eye pattern and shedding of large-scale eddies. The appearance of the cat's eye pattern is attributed to K-H instability of the shear layer, Fig. 5 and 6. Every roll is associated with a near-wall secondary eddy that might have some influence on downstream development. The K-H rolls convect downstream at a speed of $0.43 U_\infty$ and break down near the mean reattachment at $x = 8.99$ for $Tu = 1.2\%$. Further, the vortices convect at a speed of $0.45 U_\infty$, and the shear layer reattaches earlier at $x = 7.0$ for a turbulence level of 3.3%. However, the coherent eddies revealing the cat's eye patterns disappear, and the shear layer undergoes breakdown to abundant small-scale structures earlier near $x = 5.5$ with an elevated turbulence level, $Tu = 10.3\%$, Fig. 7. Thus, the change of flow features with increasing *fst* is manifested in the vorticity dynamics of an LSB.

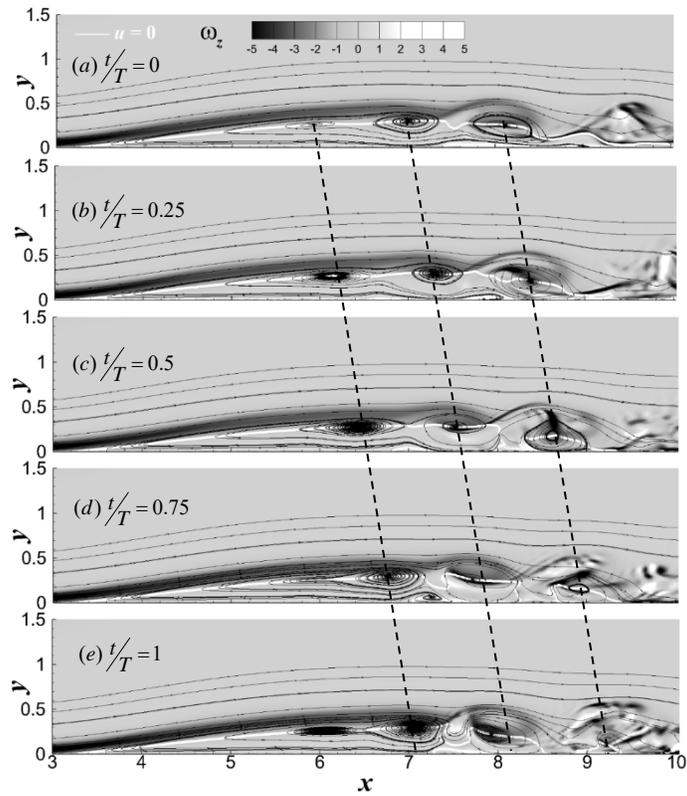

FIG. 5: Instantaneous snapshots of the spanwise vorticity over a vortex-shedding cycle with time-period $T=10$ at $z=1.5$ for $Tu=1.2\%$.

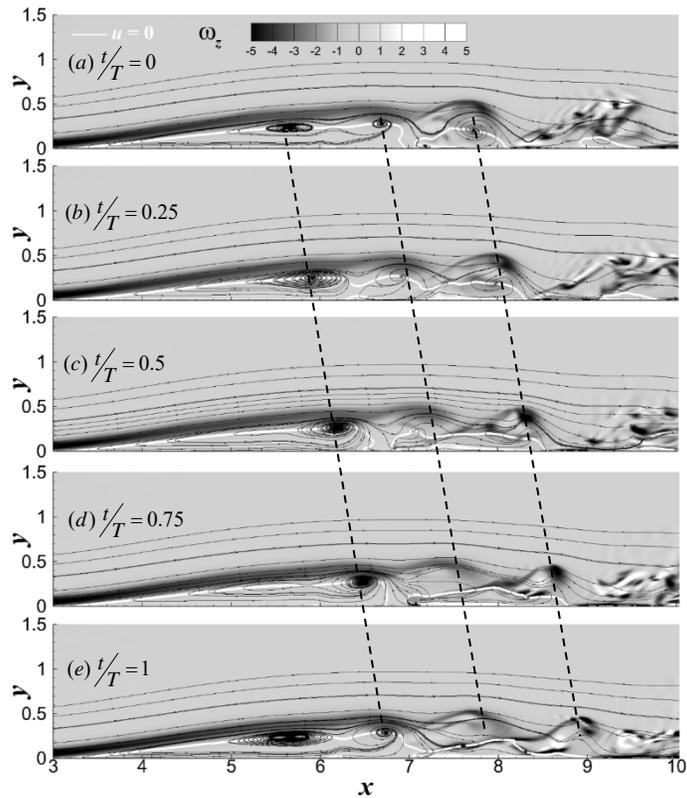

FIG. 6: Instantaneous snapshots of the spanwise vorticity over a vortex-shedding cycle with time-period $T=8.5$ at $z=1.5$ for $Tu=3.3\%$.

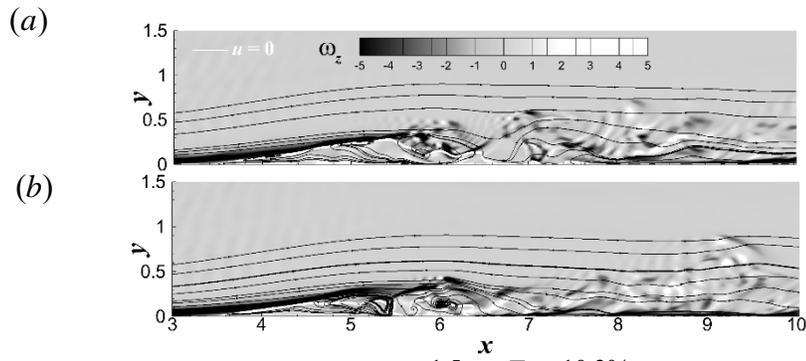

FIG 7: Instantaneous snapshots of the spanwise vorticity at $z=1.5$ for $Tu=10.3\%$ with time interval $t=2$ between two snapshots (a) and (b) in non-dimensional time unit.

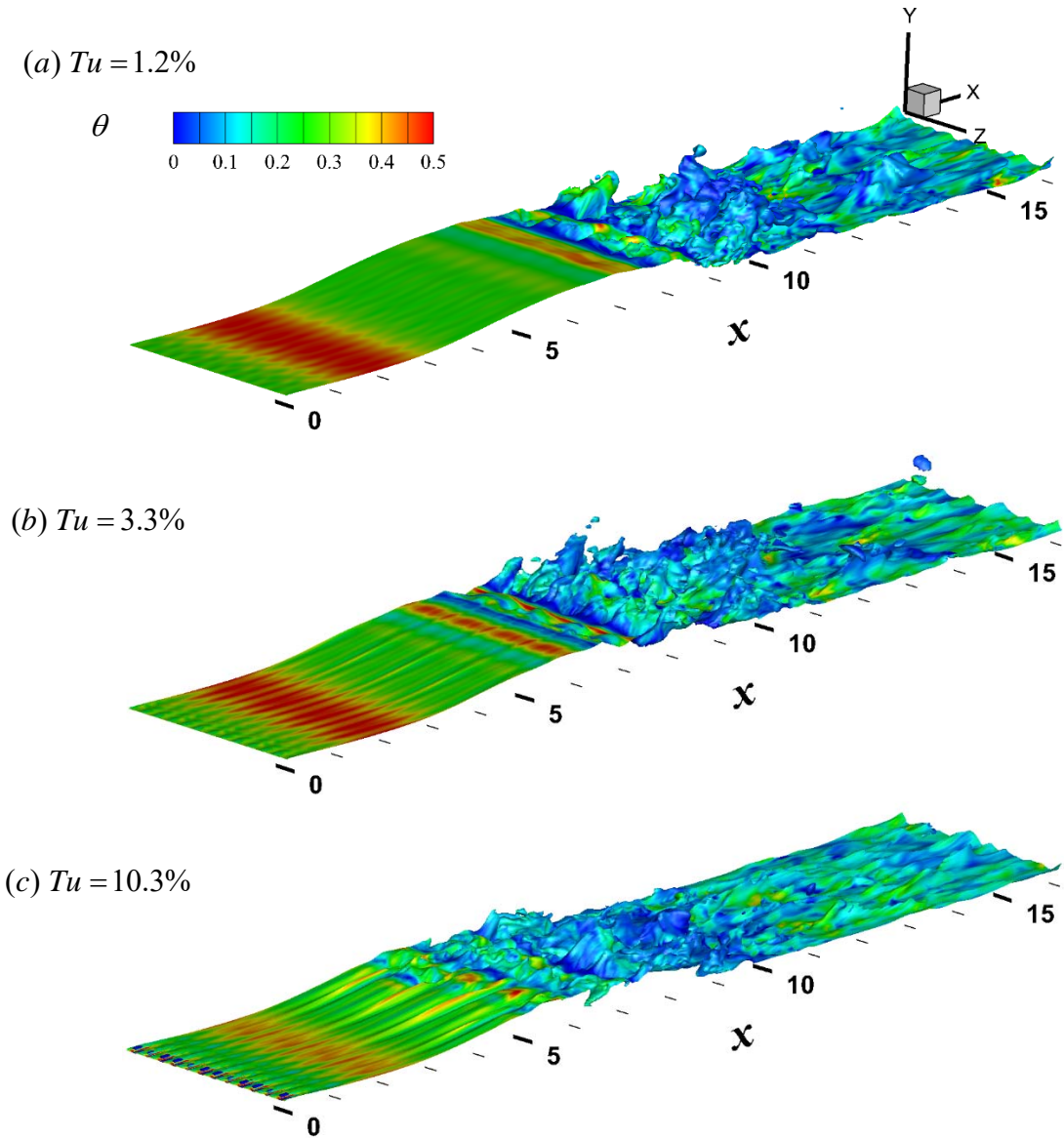

FIG. 8: Iso-surface of streamwise velocity $u=0.5$, colored by temperature for (a) $Tu=1.2\%$, (b) $Tu=3.3\%$ and (c) $Tu=10.3\%$.

The iso-surface of instantaneous streamwise velocity coloured with temperature provides a qualitative idea of the flow and thermal fields for varying levels of *fst* in Figs. 8(a-c). At the lowest *fst* level of 1.2%, a consistently heated region before separation is observed, while the heated region is progressively diluted with increasing turbulence level. Furthermore, the thermal field of the shear layer does not alter significantly before the breakdown near the reattachment at $x = 8.99$, where the three-dimensional motions are apparent downstream. This promotes significant mixing of colder free-stream fluid with the relatively hotter near-wall fluid, which can be visualized by the color scale, Fig 8(a). Thus, the breakdown of the bubble leads to an augmentation of turbulence level and the associated wall-heat transfer. It is noteworthy is that increased free-stream perturbations are eventually manifested to the streamwise streaks, Fig 8(b-c). At *fst* of 3.3%, three distinct streamwise streaks can be visualized in the dead air region; however, these streaks barely change the thermal field except by introducing some spanwise periodicity till the breakdown. The formation of stronger longitudinal streaks at an elevated *fst* of 10.3% significantly alters the temperature field even before the separation, Fig 8(c). In this case, spanwise periodic hot and cold fluid can be observed. The appearance of three-dimensional motions occurs earlier, where the flow and the thermal field are highly correlated. However, the thermal field before separation and in the dead air region have a much weaker influence of *fst*, except at a high level of *fst*. In brief, the transition of the shear layer, the development of turbulent flow downstream, and the surface heat transfer are strongly influenced by the free-stream turbulence.

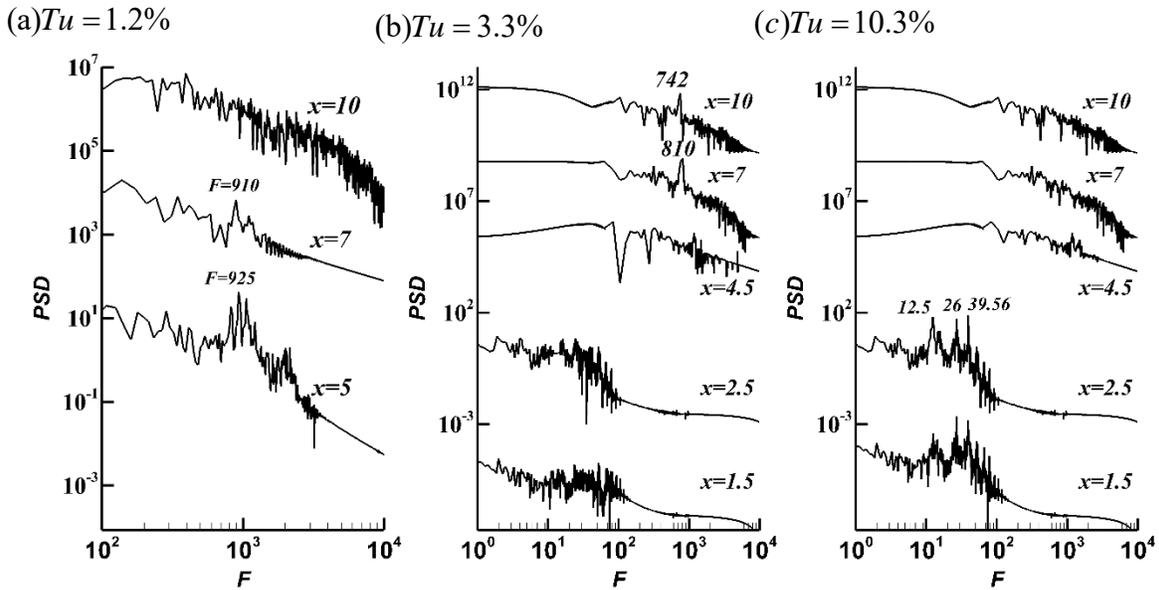

FIG 9: Power spectral density at different streamwise locations, evaluated at $y = 0.1, z = 1.5$, for (a) $Tu = 1.2\%$, (b) $Tu = 3.3\%$ and (c) $Tu = 10.3\%$.

The power spectra evaluated from the streamwise velocity history at several streamwise locations are presented in Fig. 9 (a-c) for the turbulence intensity of 1.2%, 3.3% and 10.3% to gain insights into the transition mechanism. The frequency $f$ is made dimensionless as $F = \left(2\pi v f / U_\infty^2\right) \times 10^6$, where the power spectral density (PSD) is arbitrarily normalized.[19] For a relatively low *fst* of 1.2%, a band of normalized frequency $F$ in the range 800-1200 is observed at $x = 5$ and $x = 7$ in Fig. 9(a), depicting the amplification of selective frequencies typically observed in an LSB. Further, a dominant peak and it's harmonic close to 900 and 1100 are evident in the second half of the bubble. These values of shedding frequency lie in the range of the K-H mode.[13, 30, 67] What is worthwhile is that broadband of fluctuations with normalized frequencies less than 90 are apparent at $x = 1.5$ and $x = 2.5$ for the *fst* level of 3.3% in Fig 9(b). This indicates the development of weaker longitudinal streaks via Klebanoff mode as discussed in Fig. 8 for this case. Further, the peaks at $F = 742$ and 810 seen beyond $x = 4.5$, i.e., in the second half of the bubble, confirm the appearance of K-H mode. Thus, these two modes, i.e., the K-H and Klebanoff mode, might exist and contribute together to the transition of an LSB at a moderate level of *fst*, which will be further addressed while presenting Fig. 11. At a high *fst* of 10.3%, a low-frequency band is noticed with multiple peaks at $F = 12.5, 26$ and $39.6$ prior to the separation (at $x = 1.5$ and $x = 2.5$) in Fig. 9(c). This illustrates the presence of the Klebanoff mode with low-frequency perturbations of significant amplitude, where $F$ being less the 40.[19] Further, a wide band of frequencies is observed for $x > 4.5$, indicating breakdown and downstream evolution of turbulence. Thus, the longitudinal low-frequency steaks, which appeared in the first half of the bubble via Klebanoff mode for a relatively high *fst*, interact with the free-stream perturbations and eventually manifest to abundant small-scale-eddies.

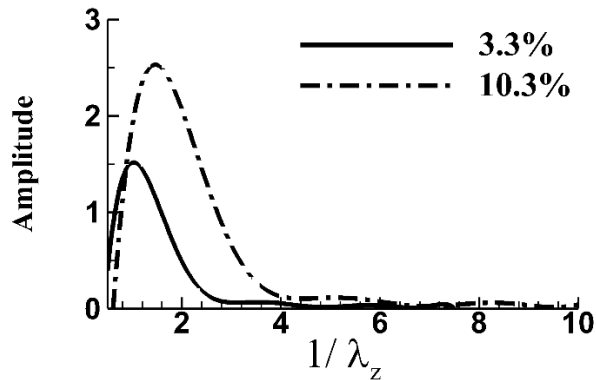

FIG. 10: Wave number spectrum of streamwise velocity in *z*-direction, evaluated at $x = 5$, *y*=0.1 for *fst* level of 3.3% (solid line) and 10.3% (dash-dot line).

To further characterize the appearance of longitudinal streaks in the first half of the bubble, the spanwise wavenumber spectrum is evaluated after autocorrelation of streamwise velocity at turbulence intensities of 3.3% and 10.3% in Fig 10. The excitation of the boundary layer by *fst* manifesting low-frequency streamwise streaks via Klebanoff modes was reported earlier[20, 30, 31, 68], where the spanwise wavenumber of $\lambda_z$ was observed in the range of $2.65\delta - 4\delta$. Figure 10 here illustrates the spanwise spacing of these streamwise streaks evaluated at a location of $x = 5$ lies in the range of $\lambda_z = 2.45\delta - 3.125\delta$, where $\delta$ is the local shear layer thickness assessed from the velocity profile. Thus, the present study corroborates the earlier observations confirming the appearance of streamwise streaks via Klebanoff modes in a separated boundary layer at moderate to high levels of *fst*.

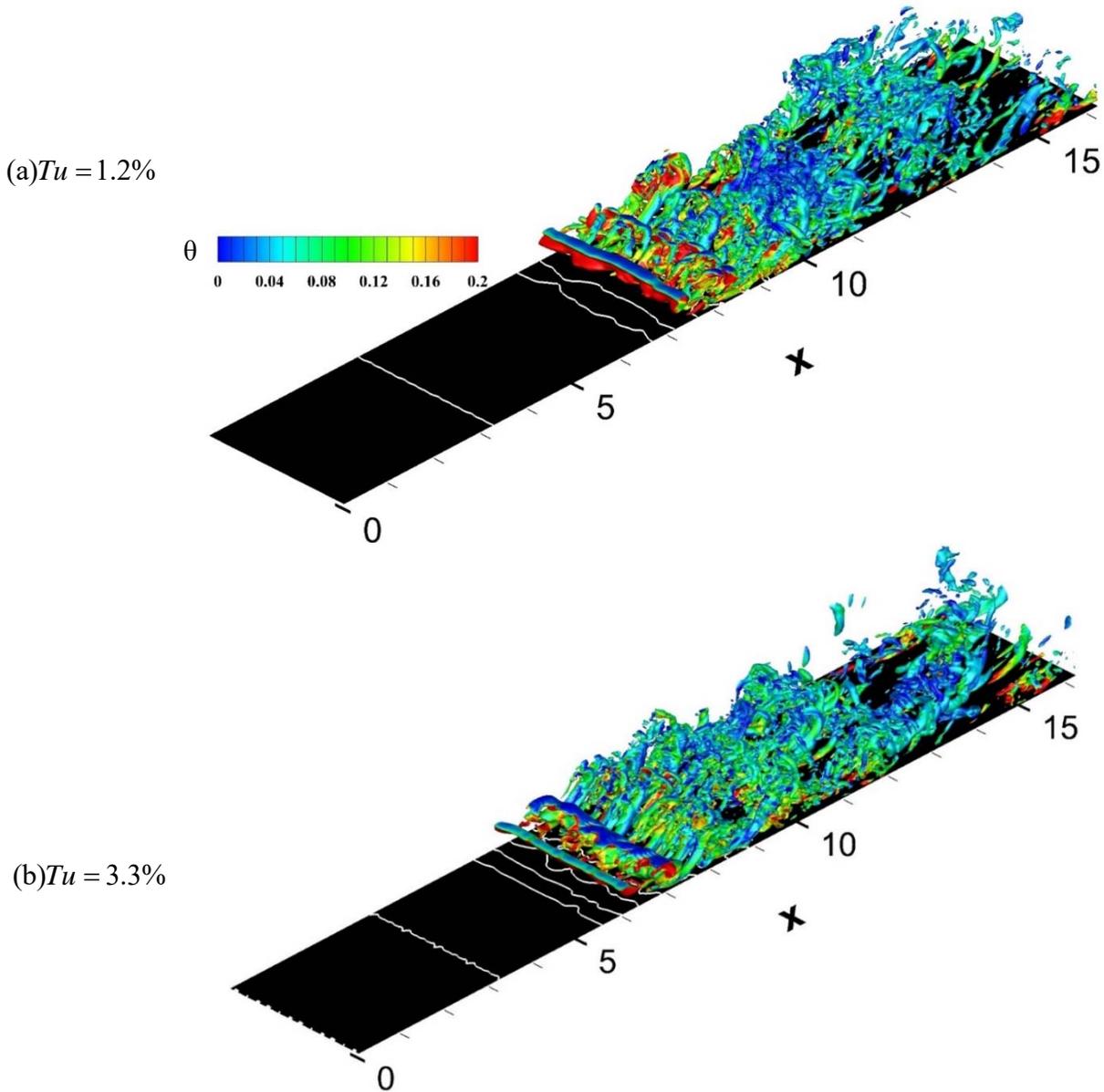

(a) $Tu = 1.2\%$

(b) $Tu = 3.3\%$

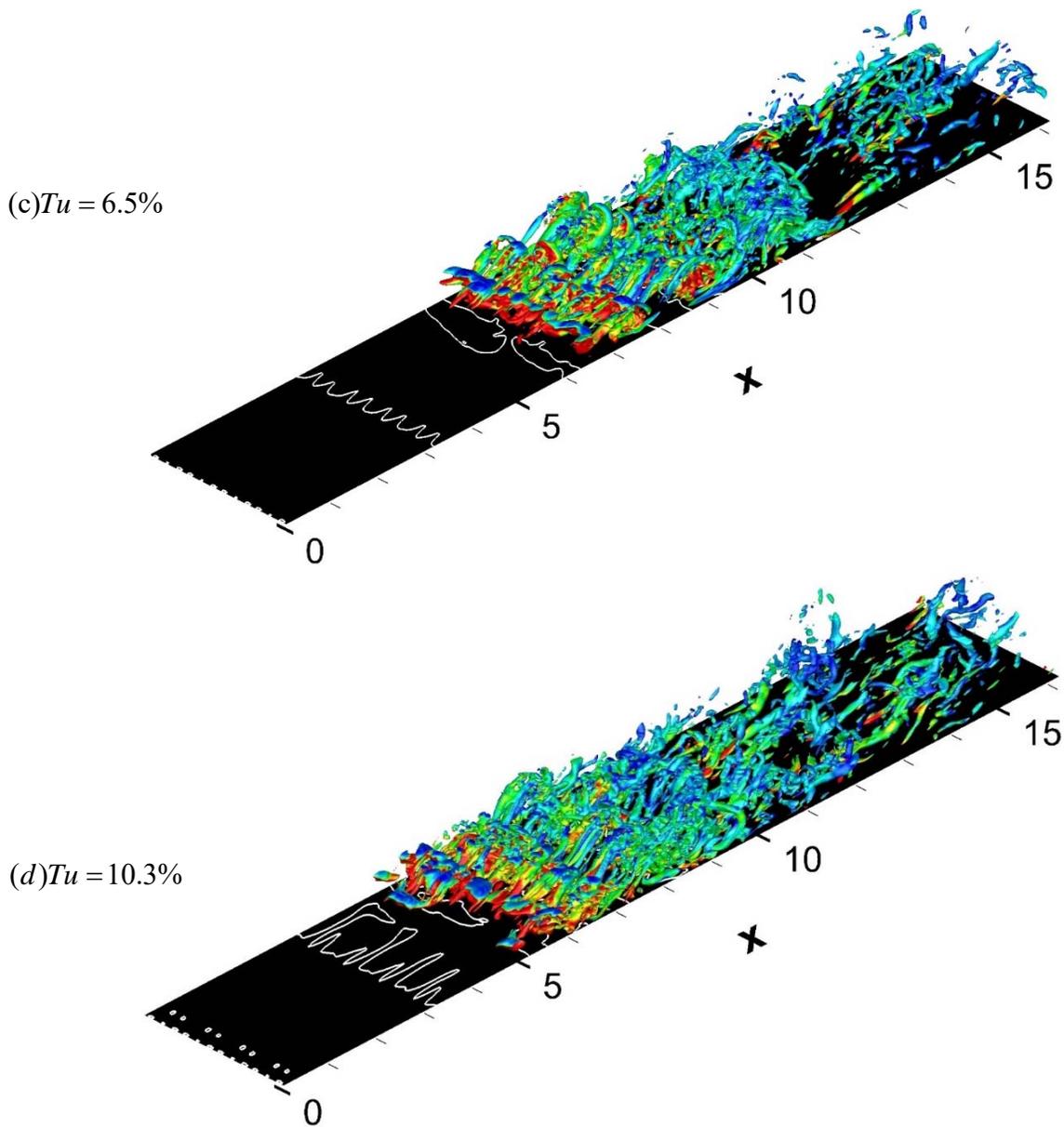

FIG. 11: Iso-surface of $Q = 2$, colored by temperature for four *fst* levels: (a) $Tu = 1.2\%$, (b) $Tu = 3.3\%$, (c) $Tu = 6.5\%$ and (d) $Tu = 10.3\%$.

Figure 11 shows the evolved flow structures using the iso-surface of the Q-criterion, coloured with their associated temperature. Contours of $\omega_z = 0$ marked by the white lines on the surface indicate either the separation or the reattachment lines. The mean separation point is nearly invariant at $x \sim 3.25$; however, the reattachment moves upstream with increasing *fst*. At $Tu = 1.2\%$, the separation line is hardly perturbed by the free-stream disturbances, while the appearance of multiple reattachment lines and their spanwise distortion indicate the existence of secondary eddies with increasing levels of fluctuations, Fig. 11(a). Here, the K-H roll appears, breaking down into small-scales energetic eddies downstream via the formation of hairpins near the reattachment at $x \sim 8.99$. Advection of hairpins and lifting their head away from the surface enhance scalar mixing and thus surface normal heat flux as indicated by associated temperature.[69] What is interesting is that the separation line is modulated in the spanwise direction as *fst* is increased to 3.3%, Fig. 11(b). This is attributed to the streamwise Klebanoff streaks, having a spanwise wavenumber of $\lambda_z = 2.65\delta - 4\delta$, as explained earlier. Further, the K-H roll is also resolved with a considerable spanwise distortion of the reattachment line near $x = 7$. This indicates that both K-H and Klebanoff modes coexist, contributing to the transition of the shear layer. For elevated *fst*, such as $Tu = 6.5\%$ and $10.3\%$, the separated line is severely modulated by streamwise streaks, and there are no traces of K-H rolls, Figs. 11(c-d). Thus, the transition of the shear layer is clearly driven by the Klebanoff modes that result in the highly three-dimensional flows with an abundance of small-scale eddies earlier.

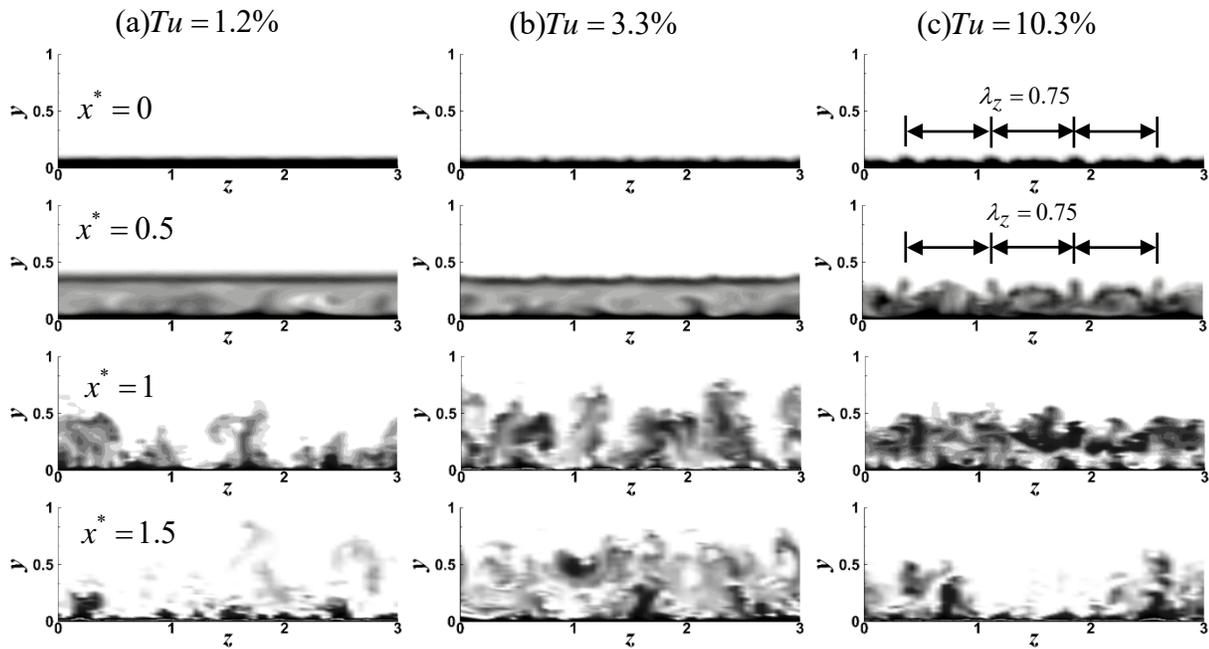

FIG. 12: Instantaneous contours of temperature in y-z planes at different $x^*$ locations, where $x^* = (x - x_s)/l_b$ for three *fst* levels (a) $Tu = 1.16\%$, (b) $Tu = 3.3\%$ and (c) $Tu = 10.3\%$.

Instantaneous contours of temperature at different cross-stream planes are presented in Fig. 12 for varying levels of *fst*. Here the streamwise coordinates are normalized as $x^* = (x - x_s)/l_b$. Thus, $x^* = 0$ and $x^* = 1$ represent the separation point and reattachment point, respectively. At $x^* = 0$, the thermal layer is seen undisturbed for a low value of imposed *fst* of 1.2%, while some weak ripples in the thermal boundary layer can be seen to develop with an increase of *fst* to 3.3%. These spanwise undulations in the thermal boundary layer become prominent at a high *fst* of 10.3%, and they bear a distinct spanwise symmetry of wavelength of $\lambda_z = 0.75$, which is the same as the Klebanoff streaks. At $x^* = 0.5$, an almost two-dimensional heated shear layer can be seen to develop over the bubble for $Tu = 3.3\%$. However, the periodic heated plumes, preserving a spanwise wavelength of $\lambda_z = 0.75$, evolve from the shear layer at $Tu = 10.3\%$. These plumes are attributed to the Klebanoff streaks, leading to scaler mixing and high heat transfer earlier. At $x^* = 1$, growing Marcum-like large-scale structures with a spanwise orientation are evident for $Tu \leq 3.3\%$. For $Tu = 10.3\%$, the appearance of abundant small-scale eddies and enhanced thermal mixing are observed from $x^* = 1$ and beyond, where the spanwise symmetry is completely lost.

## C. Mean Flow and Thermal Characteristics

The time Figure 13 compares time-averaged $u_m$, $u_{rms}$ and $\theta_{rms}$ profiles at different streamwise locations for varying *fst* levels. An inflectional velocity profile, which is invariant with *fst*, is observed at $x^* = 0$, while they deviate for varying *fst* within the separation bubble, depicting larger flow reversal for a lower *fst*, Fig. 13a. The shear layer relaxes to a turbulent boundary downstream of reattachment. The evolution of $u_{rms}$ profiles illustrates that amplification of $u_{rms}$ occurs in the second half of the bubble at a faster growth rate, followed by the saturation of turbulence after $x^* = 1$ for $Tu \leq 3.3\%$, Fig. 13(b). This is attributed to the formation of large-scale eddies and their breakdown via KH instability. Thus, turbulence augmentation and production are concentrated along the shear layer, i.e., away from the wall. It is worthwhile that significant growth of $u_{rms}$ is observed in the first half of the shear layer, where the maximum $u_{rms} = 0.035$ even at the point of separation for $Tu = 10.3\%$, while a slow decay of $u_{rms}$ is seen from $x^* = 0.5$. Thus, the shear layer appears pre-transitional, where a transient growth of perturbations mostly prevails for a high *fst* of 10.3%, which is attributed to the transition via Klebanoff mode. Amplification of $\theta_{rms}$ is similar to $u_{rms}$, illustrating an active outer layer in the first half of the bubble with a double peak nature: Inner peak is because of growing wall turbulence, and the outer peak is due to the non-linear interactions of perturbations of *k*-mode and free-stream at a higher *fst* of 10.3%, Fig 13(c). The outer peak attains maximum near the separation point, then decays, where it almost dies down after $x^* = 0.7$. The inner peak progressively grows in the second half, becomes maximum near $x^* = 0.8$, and then slowly decays, illustrating the saturation of turbulence irrespective of *fst* levels. Thus, it is worthwhile that $\theta_{rms}$ becomes an almost self-similar turbulent profile after the breakdown.

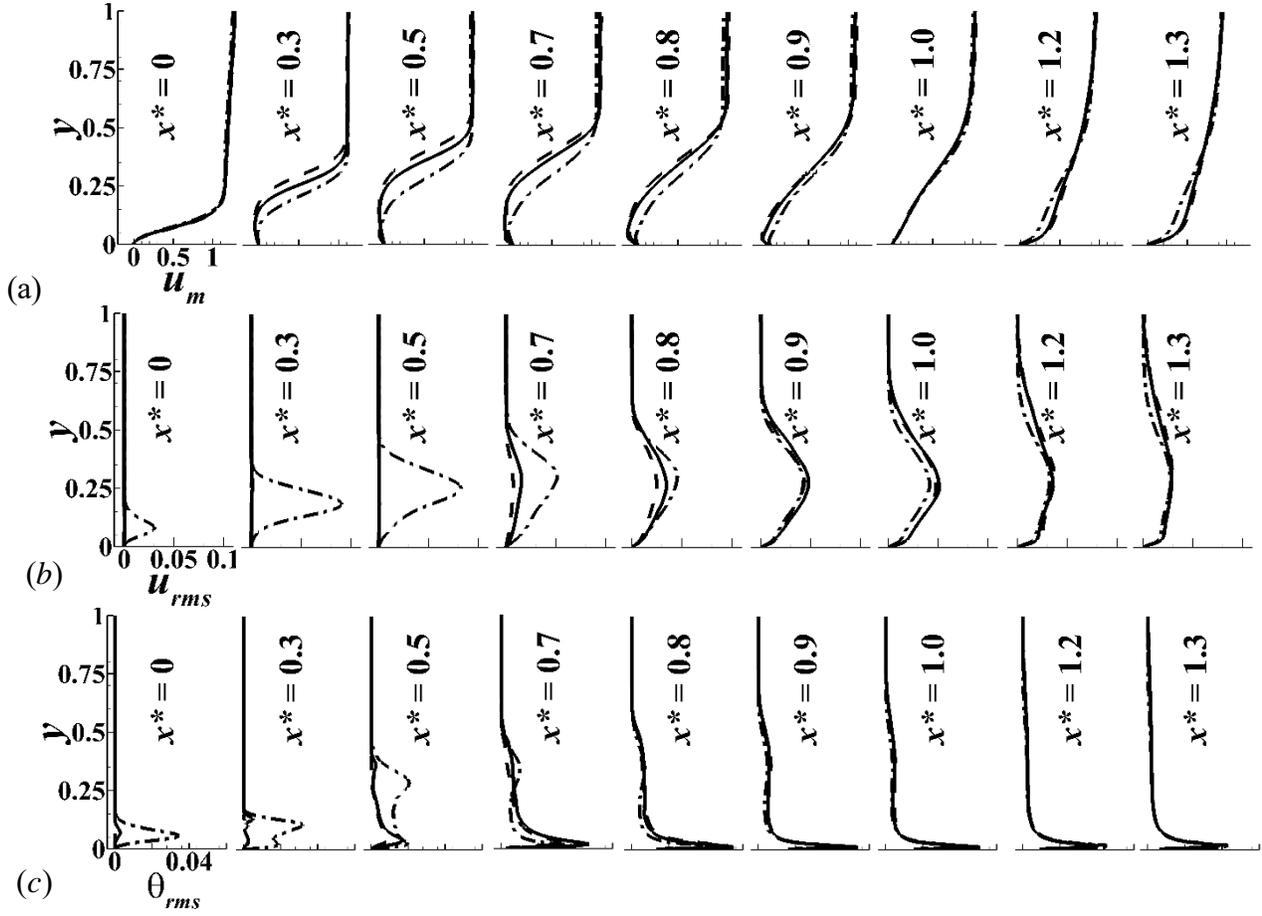

FIG. 13: Time- and span-averaged (a) mean streamwise velocity, (b) root mean square streamwise velocity and (c) temperature at different streamwise stations along the length of the bubble, $x^* = (x-x_s)/l_b$, for three *fst* levels 1.2% (dashed line), 3.3% (solid line) and 10.3% (dash-dot line).

Furthermore, the cross-stream profiles of $v_{rms}$ and $w_{rms}$ are depicted in Fig. 14 to illustrate the growth of perturbations along the shear layer and the development of three-dimensional motions. The wall-normal fluctuation, $v_{rms}$, remains small in the first half of the bubble, except for the highest *fst* level. The value of $v_{rms}$ becomes almost 8% of free-stream velocity at $x^* = 0.5$ for *fst* of 10.3%, indicating a pre-transitional shear layer in the first half. It is noted that $v_{rms}$ is associated with the wall normal heat flux. For the second half of the bubble $v_{rms}$ is higher with a lower *fst* level due to the presence of large-scale energetic K-H rolls compared to the small-scale vortices present in the region for $Tu = 10.3\%$. Downstream of the bubble ($x^* > 1$), the profiles gradually become independent of *fst*, indicating the evolution of a saturated turbulent boundary layer. The evolution of significant three-dimensional motions can also be appreciated from the profiles of $w_{rms}$. It becomes significant in the second half of the bubble for lower *fst* levels, while in even the first half of the bubble for an elevated level of *fst* of 10.3%.

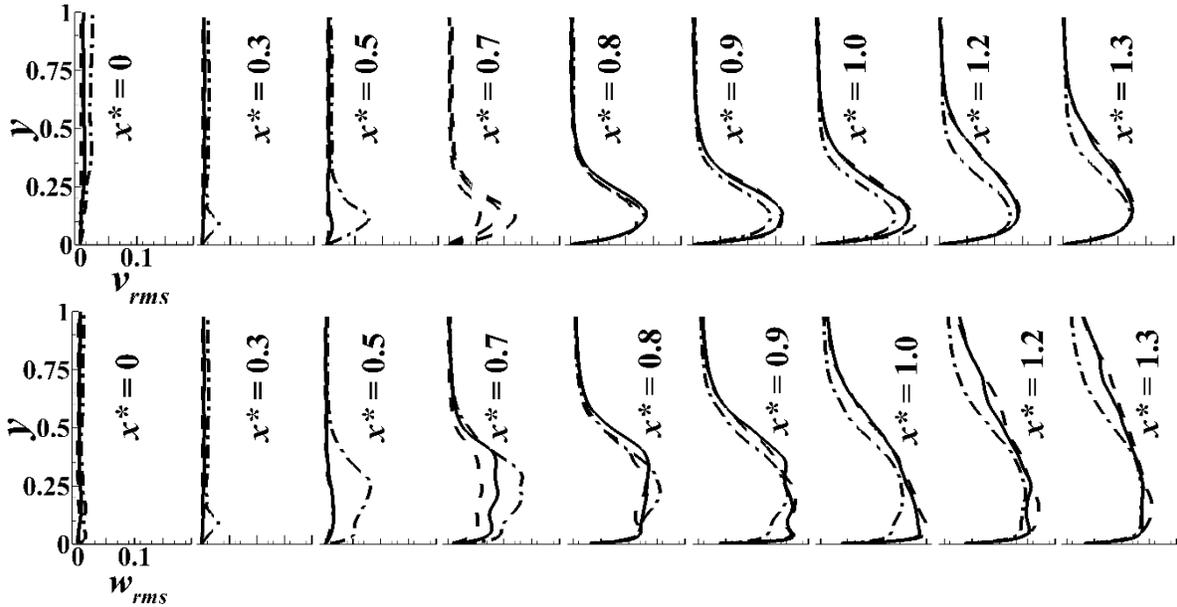

FIG. 14: Time- and span-averaged root-mean square (a) wall-normal velocity and (b) spanwise velocity profiles at different streamwise stations along the bubble, $x^* = (x - x_s)/l_b$, for three *fst* levels 1.2% (dashed line), 3.3% (solid line) and 10.3% (dash-dot line).

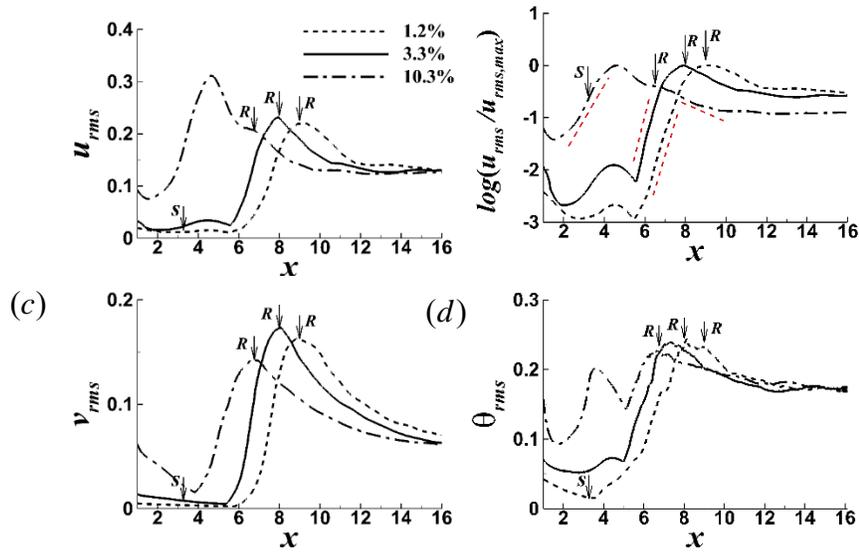

FIG. 15: Streamwise distribution of (a) the local maximum root-mean square streamwise velocity, (b) growth rate of the root mean square streamwise velocity (c) the local maximum of root-mean square wall-normal velocity and (c) growth rate of the root mean square temperature for three *fst* levels 1.2% (dashed line), 3.3% (solid line) and 10.3% (dash-dot line).

The evolution of local maximum velocity fluctuations (rms-values) along the shear layer are depicted for varying *fst* levels in Fig. 15. For *fst* levels of 1.2 and 3.3%, an exponential growth of $u_{rms}$ is observed in the second half of the bubble, where the peak appears near the reattachment followed by a decay of fluctuations, illustrating saturation of turbulence, Fig 15(a).

However, an algebraic growth of $u_{rms}$ is evident even before separation at *fst* of 10.3%, while the peak appears in the first-half of the bubble with a slow decay thereafter. The growth rate of streamwise velocity fluctuations is projected as $\log(u_{rms}/u_{rms_{max}})$ in Fig 15(b). For the *fst* levels of 1.2% and 3.3%, $(u_{rms})_{max}$ indicate a growth rate 1.16 and 1.43 respectively in the second half of the bubble and reach a peak at the reattachment, illustrating transition and subsequent evolution of three-dimensional motions primarily via K-H instability. Spalart and Strelets[18] observed a growth rate between 1 to 4 in the transition region of an LSB over a flat plate. For the *fst* level of 10.3%, $(u_{rms})_{max}$ grows linearly with a rate of 1.02 in the first-half of the bubble and attains a peak at 27% of bubble length. The saturation of turbulence thereafter is reflected by a growth rate of -0.61. However, $(v_{rms})_{max}$ exhibits a slow decay in the dead air region, followed by a sharp rise in the second half of the bubble, and becomes maximum near the reattachment for all *fst* levels, Fig 15(c). Thus, the augmentation of the cross-stream turbulence might be associated with the breakdown of the separation bubble. Furthermore, the maximum enhancement of $v_{rms}$ (value being 0.17) occurring at a $Tu = 3.3\%$ might be attributed to the interactions of the Klebanoff streaks and the large-scale eddies via K-H instability. Augmentation of temperature fluctuations ($\theta_{rms}$) can be correlated both with $u_{rms}$ and $v_{rms}$. At Tu = 1.2% and 3.3%, $\theta_{rms}$ remains almost invariant in the dead-air region, increases in the second-half and becomes maximum near the reattachment, Fig 15(d). At *fst* level of 10.3%, perturbations due to the prevalence of the streamwise streaks penetrate the separated layer, attributing to a local peak in $\theta_{rms}$ at around 25% of bubble length, followed by decay and growth, again attaining a second peak near the reattachment.

The streamwise evolutions of turbulent kinetic energy $\left(TKE = \frac{1}{2}\overline{u_i' u_i'}\right)$ and production $\left(PKE = -\overline{u_i' u_j'}\frac{\partial \overline{u_i}}{\partial x_j}\right)$ along the shear layer are presented in Fig. 16. For $Tu = 1.2\%$ and $3.3\%$, the turbulence generation occurs primarily in the second half of the bubble, which bears a resemblance to the growth of $u_{rms}$, and qualitatively indicates the onset of transition. An exponential decay of TKE thereafter reflects the evolution of a canonical boundary layer downstream, where the turbulence level becomes 0.011. Furthermore, the increase of PKE in the second-half of the bubble is attributed to a combined effect of a local concentration of vorticity via convection of K-H rolls and a high level of turbulence. On the contrary, evolutions of both TKE and PKE increase from the beginning for $Tu = 6.5\%$ and $10.3\%$. A peak appears in the first-half of the bubble, followed by decay, and increases again, resolving another local peak near the reattachment. The manifestation is due to the longitudinal streaks, their interactions with the shear layer and the concentration of velocity gradient. Further, the decay of TKE and hence PKE following the first peak is attributed to the local flow acceleration along the shear layer.

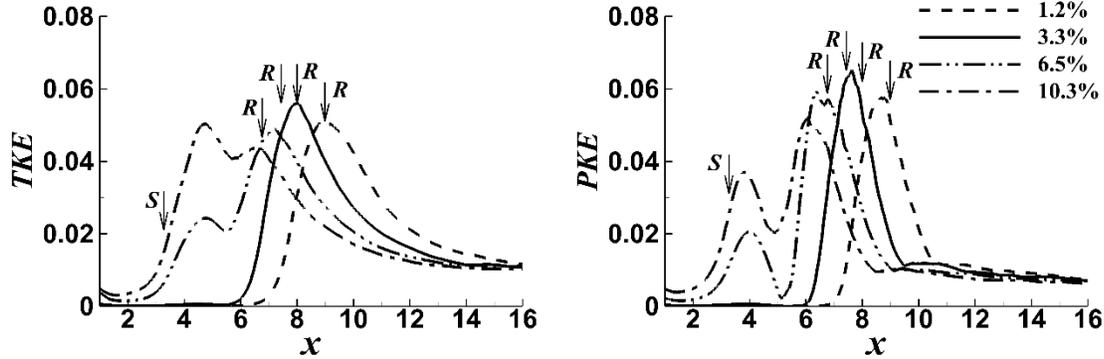

FIG. 16: Streamwise distribution of the local maximum of (a) TKE and (b) PKE, for four *fst* levels: 1.2% (dashed line), 3.3% (solid line), 6.5% (dash-dot-dot line) and 10.3% (dash-dot line).

The variations of time-averaged Nusselt ($Nu$) number in the streamwise direction is presented in Fig. 17 for different *fst* levels to appreciate the overall surface heat transfer. After separation, a plateau in the distribution of Nu number is observed in the dead air region, where the surface normal heat flux is almost unaffected by the imposed *fst*. The Nu number begins to increase and differ significantly in the transition regime, illustrating the influence of the transition mechanism in the second-half of the bubble with an increase of *fst*. Although the peak in Nu number appears near the reattachment irrespective of *fst*; the maximum value decreases with the increase of *fst* level. It is worthwhile to state that the present simulation resolved two distinct roots, which govern the momentum transfer and surface heat flux with varying imposed disturbances as (i) evolution of large-scale vortices via inviscid instability in the second half and subsequent breakdown, (ii) formation of longitudinal streaks via K-modes prior to separation and evolution of small-scale eddies due to instability of these streaks. Thus, it can be appreciated that the coherent energetic eddies in the shear layer at relatively low *fst* levels promote advection and bulk mixing of the fluid, attributing to a significant increase in heat flux. Although the heat flux increases with an increase of *fst* and associated small-scale turbulent mixing, it is not necessarily as significant as by the K-H rolls. Thus, the surface heat flux here depends on the delicate balance between large and small-scale activities.

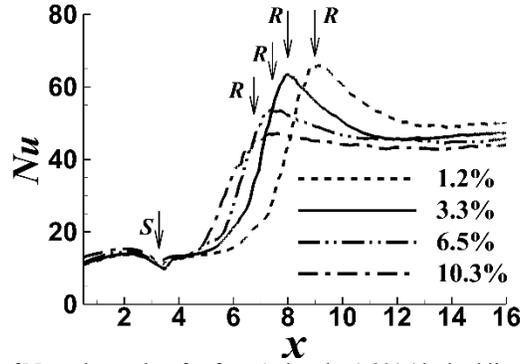

FIG. 17: Streamwise variation of Nusselt number for four *fst* levels: 1.2% (dashed line), 3.3% (solid line), 6.5% (dash-dot-dot line) and 10.3% (dash-dot line).

The streamwise and wall-normal turbulent heat flux profiles ($\overline{u'\theta'}$ and $-\overline{v'\theta'}$) are illustrated in Fig. 18 to provide further insights. The amplification of *fst* leads to a concentration of turbulent heat flux along the shear layer in the second half of the bubble. However, the heat flux (particularly $\overline{u'\theta'}$) becomes significant even in the dead-air region for $Tu = 10.3\%$, which can be attributed to the appearance of longitudinal starks via K-mode. Perturbations evolved either by K-H or by K-mode are redistributed across the shear layer, and viscous effects also become significant, resolving a double peak nature in $-\overline{v'\theta'}$ in the second half of the bubble, Fig. 18(b). The outer peak corresponds to enhanced fluctuations due to outer shear layer activities, while the inner peak resembles the viscous effects. The inner peak in $-\overline{v'\theta'}$ detects the wall heat flux, which becomes maximum near the reattachment irrespective of an imposed *fst*.

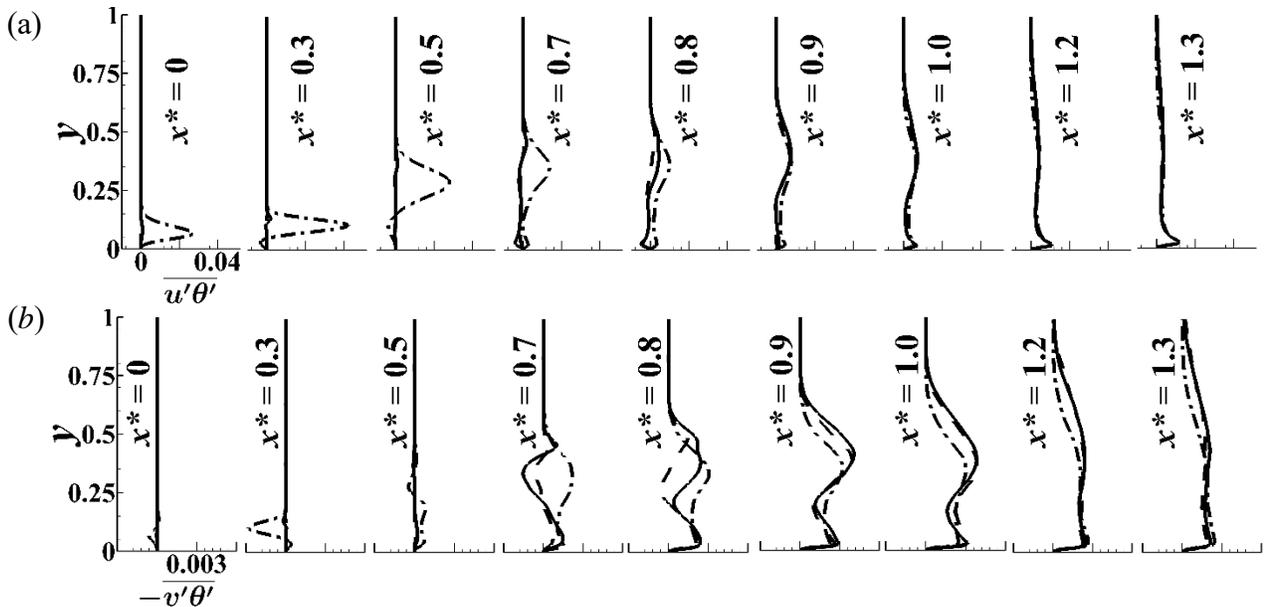

FIG. 18: Time and span averaged (a) stramwise and (b) wall normal heat flux profiles at different stations along the length of the bubble $x^* = (x-x_s)/l_b$, for three *fst* levels: 1.2% (dashed), 3.3% (solid) and 10.3% (dash-dot).

## IV. CONCLUSIONS

The manifestation of a laminar separation bubble and associated scalar transport is discussed under the influence of varying levels of free-stream turbulence using high-fidelity LES. The excitation of a shear layer changes significantly with an increase of *fst*, where the mean reattachment points shift upstream, and the separation location remains almost invariant. For a relatively low *fst* level ($Tu < 3.3\%$), the transition of an LSB occurs via inviscid instability, evolving large-scale vortices in the second half of the bubble, leading to a subsequent breakdown. The amplification of selective frequency is evident here, where the normalized frequency falls within the K-H band, which is typically observed in an LSB. Moreover, the coherent energetic eddies in the shear layer at relatively low *fst* promote advection and mixing of fluid, attributing to a significant increase in heat flux. For a relatively high *fst* ($6.5\% < Tu < 10.3\%$), the longitudinal low-frequency steaks appeared on the contrary via Klebanoff mode even prior to separation, eventually manifesting to abundant small-scale-eddies due to the instability of these streaks. Although small-scale turbulent mixing and scalar transport in the shear layer increases with *fst*, the surface normal heat flux does not necessarily as significant as by the K-H rolls. Thus, the surface heat flux depends on the delicate balance between large- and small-scale activities. Further, a mixed mode, i.e., both K-H and K-modes coexist, contributing to the flow transition at a moderate level of *fst*, lying between 3.3% and 6.5%.

An exponential growth rate of velocity fluctuations is observed in the second half of the bubble at a low level of *fst*, where the fluctuations become maximum near the reattachment, followed by saturation of turbulence. Further, the production becomes significant in the second-half, which is attributed to a combined effect of a local concentration of vorticity via convection of K-H rolls and a high level of turbulence. At a high *fst*, the shear layer appears pre-transitional via Klebanoff mode, and an algebraic growth rate of velocity fluctuations is evident here even before separation. The production increases from the beginning with the appearance of a peak in the dead-air region. The manifestation is due to the longitudinal streaks, their interactions with the shear layer, and the concentration of velocity gradient.